\documentclass[12pt,a4paper]{article}
\makeatletter
\def\eqnarray{%
\stepcounter{equation}%
\let\@currentlabel=\theequation
\global\@eqnswtrue
\global\@eqcnt\z@
\tabskip\@centering
\let\\=\@eqncr
$$\halign to \displaywidth\bgroup\@eqnsel\hskip\@centering
$\displaystyle\tabskip\z@{##}$&\global\@eqcnt\@ne
\hfil$\displaystyle{{}##{}}$\hfil
&\global\@eqcnt\tw@$\displaystyle\tabskip\z@{##}$\hfil
\tabskip\@centering&\llap{##}\tabskip\z@\cr}
\makeatother
\newcommand{\ket}[1]{{\vert{#1}\rangle}}
\newcommand{\bra}[1]{{\langle{#1}\vert}}
\newcommand{\kett}[1]{{\vert{#1}\rangle\rangle}}
\newcommand{\braa}[1]{{\langle\langle{#1}\vert}}
\newcommand{\calh}{{\cal H}}

\newcommand{\fukuso}{{\mathbf C}}
\newcommand{\futon}{{\bf N}}

\begin{document}

\title{\sl A Generalized Hamiltonian Characterizing the Interaction of 
the Two--Level Atom and both the Single Radiation Mode and External Field}
\author{
  Kazuyuki FUJII
  \thanks{E-mail address : fujii@yokohama-cu.ac.jp }\\
  Department of Mathematical Sciences\\
  Yokohama City University\\
  Yokohama, 236-0027\\
  Japan
  }
\date{}
\maketitle
%
%
%
%
\begin{abstract}
  In this paper we propose some Hamiltonian characterizing the interaction 
  of the two--level atom and both the single radiation mode and external 
  field, which might be a generalization of that of Sch{\"o}n and Cirac 
  (quant-ph/0212068). 
  We solve them in the strong coupling regime under some conditions 
  (the rotating wave approximation, resonance condition and etc), and 
  obtain unitary transformations of four types to perform Quantum Computation. 
\end{abstract}


%
%
%
%
In this paper we consider a full model of the interaction of the two--level 
atom and both the single radiation mode and external field (periodic usually), 
which might be a generalization of that of Sch{\"o}n and Cirac \cite{SCi}. 
We treat the external field as a classical one in this paper. As a general 
introduction to this topic in Quantum Optics see \cite{AE}, \cite{MSIII}, 
\cite{C-HDG}. 
Our model is deeply related to the (quantum computational) models 
\begin{itemize}
\item[(i)]\quad trapped ions with the Coulomb interaction, 
\item[(ii)]\quad trapped ions with the photon interaction (Cavity QED). 
\end{itemize}
In our model we are especially interested in the strong coupling regime, 
\cite{MFr2}, \cite{KF2}, \cite{KF3}. The motivation is a recent interesting 
experiment, \cite{NPT}. See \cite{C-HDG} and \cite{MFr3} as a general 
introduction\footnote{\cite{C-HDG} is thick but strongly recommended}. 

In \cite{MFr2} and \cite{KF2} we treated the strong coupling regime of 
the interaction model of the two--level atom and the single radiation mode, 
and have given some explicit solutions under the resonance conditions and 
rotating wave approximations. 

On the other hand we want to add some external field (like Laser one) to 
the above model which will make the model more realistic (for example in 
Quantum Computation). Therefore we propose the full model. 

We would like to solve our model in the strong coupling regime. 
Especially we want to show the existence of Rabi oscillations in this regime 
because the real purpose of a series of study (\cite{KF2}, \cite{KF3}, 
\cite{KF4}) is an application to Quantum Computation (see \cite{KF1} 
as a brief introduction to it). 

We can show the Rabi oscillations and obtain Rabi frequencies in this regime 
if the external field is constant. If it is not constant then the situation 
becomes extremely difficult.

\par \vspace{5mm}
Let $\{\sigma_{1}, \sigma_{2}, \sigma_{3}\}$ be Pauli matrices and 
${\bf 1}_{2}$ a unit matrix : 
\begin{equation}
\sigma_{1} = 
\left(
  \begin{array}{cc}
    0& 1 \\
    1& 0
  \end{array}
\right), \quad 
\sigma_{2} = 
\left(
  \begin{array}{cc}
    0& -i \\
    i& 0
  \end{array}
\right), \quad 
\sigma_{3} = 
\left(
  \begin{array}{cc}
    1& 0 \\
    0& -1
  \end{array}
\right), \quad 
{\bf 1}_{2} = 
\left(
  \begin{array}{cc}
    1& 0 \\
    0& 1
  \end{array}
\right), 
\end{equation}
and 
$\sigma_{+} = (1/2)(\sigma_{1}+i\sigma_{1})$, 
$\sigma_{-} = (1/2)(\sigma_{1}-i\sigma_{1})$. 
Let $W$ be the Walsh--Hadamard matrix 
\begin{equation}
\label{eq:2-Walsh-Hadamard}
W=\frac{1}{\sqrt{2}}
\left(
  \begin{array}{cc}
    1& 1 \\
    1& -1
  \end{array}
\right)
=W^{-1}\ , 
\end{equation}
then we can diagonalize $\sigma_{1}$ as 
$
\sigma_{1}=W\sigma_{3}W^{-1}=\sigma_{1}=W\sigma_{3}W
$
 by making use of this $W$. 
The eigenvalues of $\sigma_{1}$ is $\{1,-1\}$ with eigenvectors 
\begin{equation}
\label{eq:eigenvectors of sigma}
\ket{1}=\frac{1}{\sqrt{2}}
\left(
  \begin{array}{c}
    1 \\
    1
  \end{array}
\right), \quad 
\ket{-1}=\frac{1}{\sqrt{2}}
\left(
  \begin{array}{c}
    1 \\
    -1
  \end{array}
\right)
\quad \Longrightarrow \quad 
\ket{\lambda}=\frac{1}{\sqrt{2}}
\left(
  \begin{array}{c}
    1 \\
    \lambda
  \end{array}
\right).
\end{equation}

Let us consider an atom with $2$ energy levels $E_{0}$ and $E_{1}$ (of course 
$E_{1} > E_{0}$). 
Its Hamiltonian is in the diagonal form given as 
\begin{equation}
H_{0}=
\left(
  \begin{array}{cc}
    E_{0}& 0 \\
    0& E_{1}
  \end{array}
\right).
\end{equation}
This is rewritten as 
\begin{equation}
\label{eq:2}
H_{0}=
\frac{E_{0}+E_{1}}{2}
\left(
  \begin{array}{cc}
    1& 0 \\
    0& 1
  \end{array}
\right)-
\frac{E_{1}-E_{0}}{2}
\left(
  \begin{array}{cc}
    1& 0 \\
    0& -1
  \end{array}
\right)
\equiv \Delta_{0}{\bf 1}_{2}-\frac{\Delta}{2}\sigma_{3},
\end{equation}
where $\Delta=E_{1}-E_{0}$ is a energy difference. 
Since we usually take no interest in constant terms, we set 
\begin{equation}
\label{eq:2-energy-hamiltonian}
H_{0}=-\frac{\Delta}{2}\sigma_{3}. 
\end{equation}

We consider an atom with two energy levels which interacts with external 
(periodic) field with $g\mbox{cos}(\omega_{E}t)$. 
In the following we set $\hbar=1$ for simplicity. 
The Hamiltonian in the dipole approximation is given by 
\begin{equation}
\label{eq:2-hamiltonian}
H=H_{0}+g \mbox{cos}(\omega t)\sigma_{1}
=-\frac{\Delta}{2}\sigma_{3}+g\ \mbox{cos}(\omega_{E}t)\sigma_{1}, 
\end{equation}
where $\omega_{E}$ is the frequency of the external field, $g$ the coupling 
constant between the external field and the atom. 
We note that to solve this model without assuming the rotating wave 
approximation is not easy, see \cite{MFr1}, \cite{KF5}, \cite{BaWr}, 
\cite{SGD}. 

\par \noindent 
In the following we change the sign in the kinetic term, namely from 
$-\Delta/2$ to $\Delta/2$, to set the model for other models. 
However this is minor.

\vspace{5mm} \par 
Now we make a short review of the harmonic oscillator within our necessity.
Let $a(a^\dagger)$ be the annihilation (creation) operator of the harmonic 
oscillator.
If we set $N\equiv a^\dagger a$ (:\ number operator), then we have 
\begin{equation}
  \label{eq:basic-1}
  [N,a^\dagger]=a^\dagger\ ,\
  [N,a]=-a\ ,\
  [a^\dagger, a]=-\mathbf{1}\ .
\end{equation}
Let $\calh$ be a Fock space generated by $a$ and $a^\dagger$, and
$\{\ket{n}\vert\  n\in\futon\cup\{0\}\}$ be its basis.
The actions of $a$ and $a^\dagger$ on $\calh$ are given by
\begin{equation}
  \label{eq:basic-2}
  a\ket{n} = \sqrt{n}\ket{n-1}\ ,\
  a^{\dagger}\ket{n} = \sqrt{n+1}\ket{n+1}\ ,
  N\ket{n} = n\ket{n}
\end{equation}
where $\ket{0}$ is a normalized vacuum ($a\ket{0}=0\  {\rm and}\  
\langle{0}\vert{0}\rangle = 1$). From (\ref{eq:basic-2})
state $\ket{n}$ for $n \geq 1$ are given by
\begin{equation}
  \label{eq:basic-3}
  \ket{n} = \frac{(a^{\dagger})^{n}}{\sqrt{n!}}\ket{0}\ .
\end{equation}
These states satisfy the orthogonality and completeness conditions 
\begin{equation}
  \label{eq:basic-4}
   \langle{m}\vert{n}\rangle = \delta_{mn}\ ,\quad \sum_{n=0}^{\infty}
   \ket{n}\bra{n} = \mathbf{1}\ . 
\end{equation}
Then the displacement (coherent) operator and coherent state are defined as
\begin{equation}
  \label{eq:basic-5}
      D(z) = \mbox{e}^{za^{\dagger}- \bar{z}a}\ ;\quad 
      \ket{z} = D(z)\ket{0}\quad \mbox{for} \quad z \in \fukuso .  
\end{equation}

We consider the quantum theory of the interaction between an atom with 
two--energy levels and single radiation mode (a harmonic oscillator). 
The Hamiltonian in this case is 
\begin{equation}
\label{eq:hamiltonian-(0)}
H=\omega {\bf 1}_{2}\otimes a^{\dagger}a + 
\frac{\Delta}{2}\sigma_{3}\otimes {\bf 1} +
g\sigma_{1}\otimes (a^{\dagger}+a) 
\end{equation}
where $\omega$ is the frequency of the radiation mode, $g$ the coupling 
between the radiation field and the atom, see for example \cite{C-HDG}, 
\cite{MFr2}. 

Now it is very natural for us to include (\ref{eq:2-hamiltonian}) into 
(\ref{eq:hamiltonian-(0)}), so we present the following 

\vspace{3mm} \par \noindent
{\bf Unified Hamiltonian}
\begin{equation}
\label{eq:hamiltonian-(i)}
H=\omega{\bf 1}_{2}\otimes a^{\dagger}a + 
g_{1}\sigma_{1}\otimes (a^{\dagger}+a) + 
\frac{\Delta}{2}\sigma_{3}\otimes {\bf 1} + 
g_{2} \mbox{cos}(\omega_{E} t)\sigma_{1}\otimes {\bf 1}. 
\end{equation}
Our Hamiltonian has two coupling constants.\ 
We note that our model is deeply related to the models 
\begin{itemize}
\item[(i)]\quad trapped ions with the Coulomb interaction 
\item[(ii)]\quad trapped ions with the photon interaction (Cavity QED) 
\end{itemize}
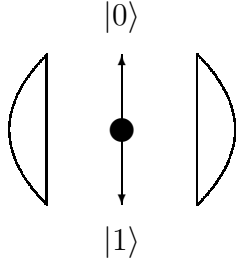
\begin{center}
\setlength{\unitlength}{1mm} 
\begin{picture}(80,40)(0,-10)
\bezier{200}(20,0)(10,10)(20,20)
\put(20,0){\line(0,1){20}}
\put(20,20){\makebox(20,10)[c]{$|0\rangle$}}
\put(30,10){\vector(0,1){10}}
\put(30,10){\vector(0,-1){10}}
\put(20,-10){\makebox(20,10)[c]{$|1\rangle$}}
\put(30,10){\circle*{3}}
\bezier{200}(40,0)(50,10)(40,20)
\put(40,0){\line(0,1){20}}
\end{picture}
\end{center}
This Hamiltonian is also related to the one presented recently by 
Sch{\"o}n and Cirac \cite{SCi}
\vspace{2mm}\par \noindent
\begin{eqnarray}
\label{eq:hamiltonian-(ii)}
H=\frac{p^2}{2m}&+& 
\omega_{0}{\bf 1}_{2}\otimes a^{\dagger}a +
g(x)\left(\sigma_{+}\otimes a+\sigma_{-}\otimes a^{\dagger}\right) +
\nonumber \\
&&\frac{\omega_0}{2}\sigma_{3}\otimes {\bf 1}+ 
\frac{\Omega}{2}
\left(
\mbox{e}^{-i\omega_{L}t}\ \sigma_{+}\otimes {\bf 1} + 
\mbox{e}^{i\omega_{L}t}\ \sigma_{-}\otimes {\bf 1}
\right). 
\end{eqnarray}
For the meaning of several constants see \cite{SCi}. 
They have assumed the rotating wave approximation (see for example 
\cite{C-HDG}) and the resonance condition, and 
use a position--dependent coupling constant $g(x)$, 
so their model is different from ours in these points.  

A comment is in order. Following \cite{SCi} 
the Hamiltonian (\ref{eq:hamiltonian-(i)}) might be modified to 
\begin{equation}
H=\frac{p^2}{2m} + \omega{\bf 1}_{2}\otimes a^{\dagger}a + 
g_{1}(x)\sigma_{1}\otimes (a^{\dagger}+a) + 
\frac{\Delta}{2}\sigma_{3}\otimes {\bf 1} + 
g_{2} \mbox{cos}(\omega_{E} t)\sigma_{1}\otimes {\bf 1}. 
\end{equation}
This model is a full generalization of (\ref{eq:hamiltonian-(ii)}), 
however we don't consider this situation in the paper. 

\vspace{3mm}\par 
We have one question : Is the Hamiltonian (\ref{eq:hamiltonian-(i)}) 
realistic or meaningful ? \quad The answer is of course yes. 
Let us show one example. 
We consider the (effective) Hamiltonian presented by NIST group 
\cite{several-1}, \cite{several-2} which were used to construct the 
controlled NOT operation (see \cite{KF1} as an introduction). 

\vspace{2mm}\par \noindent

\begin{equation}
\label{eq:hamiltonian-(iii)}
H=\omega_{0}{\bf 1}_{2}\otimes a^{\dagger}a + 
g\left(\sigma_{+}\otimes \mbox{e}^{i\eta (a^{\dagger}+a)} + 
\sigma_{-}\otimes \mbox{e}^{-i\eta (a^{\dagger}+a)}
\right) + 
\frac{\Delta}{2}\sigma_{3}\otimes {\bf 1}. 
\end{equation}

\vspace{5mm}\par
We can show that under some unitary transformation the Hamiltonian 
(\ref{eq:hamiltonian-(iii)}) can be transformed to 
(\ref{eq:hamiltonian-(i)}) with special coupling constants, 
\cite{several-3}. This is important, so we review and modify \cite{several-3}. 

We set $2A=i\eta (a^{\dagger}+a)$ for simplicity, then 
\begin{eqnarray}
&&\sigma_{+}\otimes \mbox{e}^{i\eta (a^{\dagger}+a)} + 
\sigma_{-}\otimes \mbox{e}^{-i\eta (a^{\dagger}+a)}
\nonumber \\
=&&
\left(
  \begin{array}{cc}
    0& \mbox{e}^{2A} \\
    \mbox{e}^{-2A}& 0
  \end{array}
\right)
=
\left(
  \begin{array}{cc}
    0& \mbox{e}^{A} \\
    \mbox{e}^{-A}& 0
  \end{array}
\right)
\left(
  \begin{array}{cc}
    0& 1 \\
    1& 0
  \end{array}
\right)
\left(
  \begin{array}{cc}
    0& \mbox{e}^{A} \\
    \mbox{e}^{-A}& 0  
  \end{array}
\right)  \nonumber \\
=&&
\left(
  \begin{array}{cc}
    0& \mbox{e}^{A} \\
    \mbox{e}^{-A}& 0
  \end{array}
\right)
\left\{\frac{1}{2}
\left(
  \begin{array}{cc}
    1& 1 \\
    1& -1
  \end{array}
\right)
\left(
  \begin{array}{cc}
    1& 0 \\
    0& -1
  \end{array}
\right)
\left(
  \begin{array}{cc}
    1& 1 \\
    1& -1
  \end{array}
\right)
\right\}
\left(
  \begin{array}{cc}
    0& \mbox{e}^{A} \\
    \mbox{e}^{-A}& 0  
  \end{array}
\right)  \nonumber \\
\equiv&& U(\eta)(\sigma_{3}\otimes {\bf 1})U(\eta)^{\dagger} 
\end{eqnarray}
where 
\begin{equation}
U(\eta)
=\frac{1}{\sqrt{2}}
\left(
  \begin{array}{cc}
    0& \mbox{e}^{A}   \\
    \mbox{e}^{-A}& 0
  \end{array}
\right)
\left(
  \begin{array}{cc}
    1& 1 \\
    1& -1
  \end{array}
\right)
=\frac{1}{\sqrt{2}}
\left(
  \begin{array}{cc}
    \mbox{e}^{A}& -\mbox{e}^{A}   \\
    \mbox{e}^{-A}& \mbox{e}^{-A}
  \end{array}
\right)
\end{equation}
and $\mbox{e}^{A}=D(i\eta/2)$ where $D(\beta)$ is a displacement (coherent) 
operator defined by (\ref{eq:basic-5}). 
Then it is not difficult to show 
\begin{equation}
U(\eta)^{\dagger}HU(\eta)
=\frac{\omega_0\eta^2}{4}{\bf 1}_{2}\otimes {\bf 1}+
\omega_{0}{\bf 1}_{2}\otimes a^{\dagger}a+
\frac{\omega_{0}\eta}{2}\sigma_{1}\otimes (-ia^{\dagger}+ia)+
g\sigma_{3}\otimes {\bf 1}-\frac{\Delta}{2}\sigma_{1}\otimes {\bf 1}.
\end{equation}

\par \noindent 
To remove $i$ in the term containing $a$ 
we moreover operate the unitary one 
\begin{eqnarray}
&&\left({\bf 1}_{2}\otimes \mbox{e}^{i(\pi/2) N}\right)U(\eta)^{\dagger}H
U(\eta)\left({\bf 1}_{2}\otimes \mbox{e}^{-i(\pi/2) N}\right)  \nonumber \\
=&&\frac{\omega_0\eta^2}{4}{\bf 1}_{2}\otimes {\bf 1}+
\omega_{0}{\bf 1}_{2}\otimes a^{\dagger}a+
\frac{\omega_{0}\eta}{2}\sigma_{1}\otimes (a^{\dagger}+a)+
g\sigma_{3}\otimes {\bf 1}-\frac{\Delta}{2}\sigma_{1}\otimes {\bf 1},
\end{eqnarray}
where we have used the well--known formula 
\[
\mbox{e}^{i\theta N}a\mbox{e}^{-i\theta N}=\mbox{e}^{-i\theta}a, \quad 
\mbox{e}^{i\theta N}a^{\dagger}\mbox{e}^{-i\theta N}=
\mbox{e}^{i\theta}a^{\dagger},
\]
see \cite{KF4}. Since $U(\eta)$ can be written as 
$
U(\eta)=
(\sigma_{+}\otimes \mbox{e}^{A}+\sigma_{-}\otimes \mbox{e}^{-A})
\left(W\otimes {\bf 1}\right),
$
so if we write 
\begin{equation}
T(\eta)\equiv U(\eta)({\bf 1}_{2}\otimes \mbox{e}^{-i(\pi/2) N})
=(\sigma_{+}\otimes \mbox{e}^{A}+\sigma_{-}\otimes \mbox{e}^{-A})
(W\otimes \mbox{e}^{-i(\pi/2) N}), 
\end{equation}
then we have 
\[
T(\eta)^{\dagger}HT(\eta)
=\frac{\omega_0\eta^2}{4}{\bf 1}_{2}\otimes {\bf 1}+
\omega_{0}{\bf 1}_{2}\otimes a^{\dagger}a+
\frac{\omega_{0}\eta}{2}\sigma_{1}\otimes (a^{\dagger}+a)+
g\sigma_{3}\otimes {\bf 1}-\frac{\Delta}{2}\sigma_{1}\otimes {\bf 1}. 
\]
Here we have no interest in the constant term, so we finally obtain 
\begin{equation}
H=T(\eta)
\left\{
\omega_{0}{\bf 1}_{2}\otimes a^{\dagger}a+
\frac{\omega_{0}\eta}{2}\sigma_{1}\otimes (a^{\dagger}+a)+
g\sigma_{3}\otimes {\bf 1}-\frac{\Delta}{2}\sigma_{1}\otimes {\bf 1}
\right\}T(\eta)^{\dagger}.
\end{equation}
$T(\eta)$ is just the unitary transformation required.

\vspace{5mm}\par 
At this stage we would like to make a further generalization of  
the Hamiltonian (\ref{eq:hamiltonian-(i)}) to make wide applications 
to Quantum Computation.@

Let $\{K_{+},K_{-},K_{3}\}$ and $\{J_{+},J_{-},J_{3}\}$ be a set of 
generators of unitary representations of Lie algebras $su(1,1)$ and 
$su(2)$. Then we can make the similar arguments done for the Heisenberg 
algebra $\{a^{\dagger},a,N\}$, namely (\ref{eq:basic-1}) $\sim$ 
(\ref{eq:basic-5}), see for example \cite{KF4}. 

\par 
We have considered the following three Hamiltonians in \cite{KF2} : 
\begin{eqnarray}
\mbox{(N)}\qquad H_{N}&=&\omega {\bf 1}_{2}\otimes a^{\dagger}a + 
\frac{\Delta}{2}\sigma_{3}\otimes {\bf 1} +
g\sigma_{1}\otimes (a^{\dagger}+a), \\
\mbox{(K)}\qquad H_{K}&=&\omega {\bf 1}_{2}\otimes K_{3} + 
\frac{\Delta}{2}\sigma_{3}\otimes {\bf 1}_{K} +
g\sigma_{1}\otimes (K_{+}+K_{-}), \\
\mbox{(J)}\qquad H_{J}&=&\omega {\bf 1}_{2}\otimes J_{3} + 
\frac{\Delta}{2}\sigma_{3}\otimes {\bf 1}_{J} +
g\sigma_{1}\otimes (J_{+}+J_{-}).
\end{eqnarray}

\par \noindent 
To treat these three cases at the same time we set 
\begin{equation}
\{L_{+},L_{-},L_{3}\}=
\left\{
\begin{array}{ll}
\mbox{(N)}\qquad \{a^{\dagger},a,N\}, \\
\mbox{(K)}\qquad \{K_{+},K_{-},K_{3}\}, \\
\mbox{(J)}\qquad \ \{J_{+},J_{-},J_{3}\} 
\end{array}
\right.
\end{equation}
and 
\begin{equation}
H_{L}=\omega {\bf 1}_{2}\otimes L_{3} + 
\frac{\Delta}{2}\sigma_{3}\otimes {\bf 1}_{L} +
g\sigma_{1}\otimes (L_{+}+L_{-}).
\end{equation}

\vspace{5mm}\par \noindent 
Therefore the Hamiltonian that we are looking for is 
\vspace{3mm} \par \noindent
{\bf Full Hamiltonian}
\begin{equation}
\label{eq:general-hamiltonian}
{\tilde H}_{L}
=\omega {\bf 1}_{2}\otimes L_{3} + 
g_{1}\sigma_{1}\otimes (L_{+}+L_{-}) + 
\frac{\Delta}{2}\sigma_{3}\otimes {\bf 1}_{L} + 
g_{2} \mbox{cos}(\omega_{E}t)\sigma_{1}\otimes {\bf 1}_{L}. 
\end{equation}

\par \noindent 
From now we would like to solve this Hamiltonian, especially 
in the strong coupling regime ($g_{1}\gg \Delta$). 

Let us transform (\ref{eq:general-hamiltonian}) into 
\begin{eqnarray}
{\tilde H}_{L}
&=&{\bf 1}_{2}\otimes \omega L_{3} + 
\sigma_{1}\otimes 
\left\{g_{1}(L_{+}+L_{-}) + g_{2} \mbox{cos}(\omega_{E}t){\bf 1}_{L}
\right\} + 
\frac{\Delta}{2}\sigma_{3}\otimes {\bf 1}_{L}     \nonumber \\
&\equiv& {\tilde H}_{0}+\frac{\Delta}{2}\sigma_{3}\otimes {\bf 1}_{L}. 
\end{eqnarray}
The method to solve is almost identical to \cite{KF2}, so we give only an 
outline. By making use of the Walsh--Hadamard matrix 
(\ref{eq:2-Walsh-Hadamard}) 
\begin{eqnarray}
{\tilde H}_{0}
&=&(W\otimes {\bf 1}_{L})
\left[
{\bf 1}_{2}\otimes \omega L_{3} + 
\sigma_{3}\otimes 
\left\{g_{1}(L_{+}+L_{-}) + g_{2} \mbox{cos}(\omega_{E}t){\bf 1}_{L}\right\}
\right]
(W^{-1}\otimes {\bf 1}_{L})   \nonumber \\
&=&\sum_{\lambda=\pm 1}
\left(
\ket{\lambda}\otimes \mbox{e}^{-\frac{\lambda x}{2}(L_{+}-L_{-})}
\right)
\left\{\Omega L_{3}+\lambda g_{2} \mbox{cos}(\omega_{E}t){\bf 1}_{L}\right\}
\left(
\bra{\lambda}\otimes \mbox{e}^{\frac{\lambda x}{2}(L_{+}-L_{-})}
\right) \nonumber 
\end{eqnarray}
where $\ket{\lambda}$ is the eigenvectors of $\sigma_{1}$ defined in 
(\ref{eq:eigenvectors of sigma}) and $\Omega,\ x$ are given as 
\begin{equation}
\label{eq:omega-x}
(\Omega,\ x)=
\left\{
\begin{array}{ll}
(N)\quad \omega,\quad \quad \quad \qquad \qquad \ x=2g_{1}/\omega, \\
(K)\quad \omega \sqrt{1-(2g_{1}/\omega)^{2}},\quad 
x=\mbox{tanh}^{-1}(2g_{1}/\omega), \\
(J)\quad \ \omega \sqrt{1+(2g_{1}/\omega)^{2}},\quad 
x=\mbox{tan}^{-1}(2g_{1}/\omega). 
\end{array}
\right.
\end{equation}
That is, we could diagonalize the Hamiltonian ${\tilde H}_{0}$. 
Its eigenvalues $\{E_{n}(t)\}$ and eigenvectors $\{\ket{\{\lambda, n\}}\}$ 
are given respectively 
\begin{equation}
\label{eq:Eigenvalues-Eigenvectors}
(E_{n}(t),\ \ket{\{\lambda, n\}})=
\left\{
\begin{array}{ll}
(N)\quad \Omega (-\frac{g^2}{\omega^2}+n)+
\lambda g_{2} \mbox{cos}(\omega_{E}t), \quad \ket{\lambda}\otimes 
\mbox{e}^{-\frac{\lambda x}{2}(a^{\dagger}-a)}\ket{n}, \\
(K)\quad \Omega (K+n)+\lambda g_{2} \mbox{cos}(\omega_{E}t),\quad \quad 
\ket{\lambda}\otimes 
\mbox{e}^{-\frac{\lambda x}{2}(K_{+}-K_{-})}\ket{K,n}, \\
(J)\quad \ \Omega (-J+n)+\lambda g_{2} \mbox{cos}(\omega_{E}t),\quad \ 
\ket{\lambda}\otimes 
\mbox{e}^{-\frac{\lambda x}{2}(J_{+}-J_{-})}\ket{J,n} \\
\end{array}
\right.
\end{equation}
for $\lambda=\pm 1$ and $n \in \futon \cup \{0\}$, where 
$E_{n}(t)\equiv E_{n}+\lambda g_{2} \mbox{cos}(\omega_{E}t)$. 
Then ${\tilde H}_{0}$ above can be written as 
\[
{\tilde H}_{0}
=\sum_{\lambda}\sum_{n}E_{n}(t)\ket{\{\lambda, n\}}\bra{\{\lambda, n\}}.
\]

\par \vspace{3mm} 
Next we would like to solve the following Schr{\"o}dinger equation : 
\begin{equation}
\label{eq:full-equation}
i\frac{d}{dt}\Psi={\tilde H}\Psi=\left({\tilde H}_{0}+
\frac{\Delta}{2}\sigma_{3}\otimes {\bf 1}_{L}\right)\Psi. 
\end{equation}
To solve this equation we appeal to the method of constant variation. 
First let us solve 
$
i\frac{d}{dt}\Psi={\tilde H}_{0}\Psi, 
$
which general solution is given by 
$
\label{eq:partial-solution}
  \Psi(t)=U_{0}(t)\Psi_{0}
$, 
where $\Psi_{0}$ is a constant state and 
\begin{equation}
\label{eq:Basic-Unitary}
U_{0}(t)=\sum_{\lambda}\sum_{n}
\mbox{e}^{-i\{tE_{n}+\lambda(g_{2}/\omega_{E})sin(\omega_{E}t)\}}
\ket{\{\lambda, n\}}\bra{\{\lambda, n\}}.
\end{equation}
The method of constant variation goes as follows. Changing like 
$
\Psi_{0} \longrightarrow \Psi_{0}(t),
$ 
we have 
\begin{equation}
\label{eq:sub-equation}
i\frac{d}{dt}\Psi_{0}
=\frac{\Delta}{2}{U_0}^{\dagger}(\sigma_{3}\otimes {\bf 1}_{L}){U_0}\Psi_{0} 
\equiv \frac{\Delta}{2}{\tilde H}_{F}\Psi_{0}
\end{equation}
after some algebra. We must solve this equation. ${\tilde H}_{F}$ is 
\begin{eqnarray}
{\tilde H}_{F}&=&
\sum_{\lambda,\mu}\sum_{m, n}
\mbox{e}^{it(E_m-E_n)+i(\lambda-\mu)(g_{2}/\omega_{E})sin(\omega_{E}t)} 
\bra{\{\lambda,m\}}(\sigma_{3}\otimes {\bf 1}_{L})\ket{\{\mu,n\}} \ 
\ket{\{\lambda, m\}}\bra{\{\mu, n\}} \nonumber \\
&=&
\sum_{\lambda}\sum_{m, n}
\mbox{e}^{ i\{t\Omega(m-n)+2\lambda(g_{2}/\omega_{E})sin(\omega_{E}t)\} } 
\braa{m}\mbox{e}^{{\lambda x}(L_{+}-L_{-})}\kett{n} \ 
\ket{\{\lambda, m\}}\bra{\{-\lambda, n\}}, 
\end{eqnarray}
where we have used $\bra{\lambda}\sigma_{3}=\bra{-\lambda}$ 
and $\kett{n}$ is respectively 
\[
\kett{n}=
\left\{
\begin{array}{ll}
(N)\qquad \ket{n}, \\
(K)\qquad \ket{K,n}, \\
(J)\qquad\ \ket{J,n}. 
\end{array}
\right.
\]
In the following we set for simplicity 
\begin{equation}
\label{eq:time-depend-F}
\Theta(t)\equiv g_{2}\frac{\mbox{sin}(\omega_{E}t)}{\omega_{E}}.
\end{equation}
Here we divide ${\tilde H}_{F}$ into two parts
$
{\tilde H}_{F}={{\tilde H}_{F}}^{'}+{{\tilde H}_{F}}^{''}
$
where 
\begin{eqnarray}
\label{eq:Second-Hamiltonian-1}
{{\tilde H}_{F}}^{'}&=&\sum_{\lambda}\sum_{n}\mbox{e}^{2i\lambda \Theta(t)}
\braa{n}\mbox{e}^{{\lambda x}(L_{+}-L_{-})}\kett{n} \ 
\ket{\{\lambda, n\}}\bra{\{-\lambda, n\}}, \\
\label{eq:Second-Hamiltonian-2}
{{\tilde H}_{F}}^{''}&=&\sum_{\lambda}
\sum_{\stackrel{\scriptstyle m,n}{m\ne n}}
\mbox{e}^{i\{t\Omega (m-n)+2\lambda \Theta(t) \}} 
\braa{m}\mbox{e}^{{\lambda x}(L_{+}-L_{-})}\kett{n} \ 
\ket{\{\lambda, m\}}\bra{\{-\lambda, n\}}.
\end{eqnarray}
Noting 
$
\braa{n}\mbox{e}^{x(L_{+}-L_{-})}\kett{n}=
\braa{n}\mbox{e}^{-x(L_{+}-L_{-})}\kett{n}
$
by the results in section 3 of \cite{KF2}, ${{\tilde H}_{F}}^{'}$ can be 
written as
\[
{{\tilde H}_{F}}^{'}=\sum_{n}\braa{n}\mbox{e}^{x(L_{+}-L_{-})}\kett{n} 
\left\{ 
\mbox{e}^{2i\Theta(t)}\ket{\{1, n\}}\bra{\{-1, n\}} + 
\mbox{e}^{-2i\Theta(t)}\ket{\{-1, n\}}\bra{\{1, n\}}
\right\}. 
\]
Here we want to solve the equation 
$i(d/dt)\Psi_{0}={{\tilde H}_{F}}^{'}\Psi_{0}$ completely 
, however it is not easy (see \cite{MFr1}, \cite{BaWr}, \cite{SGD} and 
Appendix). 
Therefore we make a strong assumption. Namely we consider only the constant 
external field in (\ref{eq:general-hamiltonian}) 
($\omega_{E}=0$ in (\ref{eq:time-depend-F})), so 
\begin{equation}
\label{eq:time-depend-F-special}
\Theta(t)=g_{2}t.
\end{equation}
In this case we can solve the equation 
completely\footnote{In the following the method to solve the equations is 
different from that of \cite{KF2} in which the Schr{\"o}dinger cat states were 
used. However in this case we cannot use them, so the method becomes rather 
complicated.}.

A comment is in order. For a short time span it may be not unrealistic to 
consider the above situation. Anyway, the author doesn't know whether it is 
reasonable or not. 

\par \vspace{2mm} \noindent 
Therefore we have 
\begin{equation}
\label{eq:}
{{\tilde H}_{F}}^{'}=\sum_{n}\braa{n}\mbox{e}^{x(L_{+}-L_{-})}\kett{n} 
\left\{ 
\mbox{e}^{2ig_{2}t}\ket{\{1, n\}}\bra{\{-1, n\}} + 
\mbox{e}^{-2ig_{2}t}\ket{\{-1, n\}}\bra{\{1, n\}}
\right\}. 
\end{equation}
Now it is easy to solve the equation 
$i(d/dt)\Psi_{0}={{\tilde H}_{F}}^{'}\Psi_{0}$, see Appendix. 
Next let us transform (\ref{eq:Second-Hamiltonian-2}). 
\begin{eqnarray}
\label{eq:non-diagonal-hamiltonian}
{H_{F}}^{''}&=&
\sum_{\stackrel{\scriptstyle m,n}{m\ne n}}
\mbox{e}^{it\Omega(m-n)} 
\left\{
\braa{m}\mbox{e}^{x(L_{+}-L_{-})}\kett{n}
\mbox{e}^{2ig_{2}t}|\{1, m\}\rangle \langle\{-1, n\}|\ + 
\right.      \nonumber \\
&&\left. \qquad \qquad \qquad \ 
\braa{m}\mbox{e}^{-x(L_{+}-L_{-})}\kett{n}
\mbox{e}^{-2ig_{2}t}|\{-1, m\}\rangle \langle\{1, n\}|
\right\}. 
\end{eqnarray}

\par \noindent
For simplicity in the following we set 
\begin{equation}
E_{n,\Delta}=\frac{\Delta}{2}
\braa{n}\mbox{e}^{x(L_{+}-L_{-})}\kett{n}, 
\end{equation}
then 
\begin{equation}
E_{n,\Delta}=
\left\{
\begin{array}{ll}
(N)\quad \frac{\Delta}{2}
\mbox{e}^{-\frac{\kappa^2}{2}}L_{n}\left(\kappa^{2}\right)
\quad  \mbox{where}\quad \kappa=x  \\
(K)\quad \frac{\Delta}{2}
\frac{n!}{(2K)_{n}}(1+\kappa^2)^{-K-n}
F_{n}(\kappa^2:2K)\quad \mbox{where}\quad \kappa=\mbox{sinh}(x)  \\
(J)\quad \frac{\Delta}{2}
\frac{n!}{{}_{2J}P_n}(1-\kappa^2)^{J-n}
F_{n}(\kappa^2:2J)\quad \mbox{where}\quad \kappa=\mbox{sin}(x)  
\end{array}
\right.
\end{equation}
from the results in sectin 3.1 of \cite{KF2}. 

Now let us solve (\ref{eq:sub-equation})
\[
i\frac{d}{dt}\Psi_{0}
=\frac{\Delta}{2}{\tilde H}_{F}\Psi_{0}
=\frac{\Delta}{2} ({{\tilde H}_{F}}^{'}+{{\tilde H}_{F}}^{''})\Psi_{0}.
\]
For that using the method of constant variation again 
we can set $\Psi_{0}(t)$ as 
\begin{equation}
\label{eq:full-ansatz}
\Psi_{0}(t)=\sum_{n}
\left\{
(u_{n,11}a_{n,1}+u_{n,12}a_{n,-1})\ket{\{1, n\}}+
(u_{n,21}a_{n,1}+u_{n,22}a_{n,-1})\ket{\{-1, n\}}
\right\},
\end{equation}
where from the appendix 
\begin{equation}
\label{eq:special-unitary}
U_{n}(t)=
\left(
  \begin{array}{cc}
    u_{n,11}& u_{n,12}\\
    u_{n,21}& u_{n,22}
  \end{array}
\right)=
\left(
  \begin{array}{cc}
    1& \\
     & \mbox{e}^{-2ig_{2}t}
  \end{array}
\right)
\mbox{exp}\left\{-it
\left(
  \begin{array}{cc}
    0& E_{n,\Delta} \\
    E_{n,\Delta}& -2g_{2}
  \end{array}
\right)
\right\}, 
\end{equation}
then we have a set of complicated equations with respect to 
$\{a_{n,\lambda}(t)\}$. However it is almost impossible to solve them, so 
let us make a daring assumption like \cite{KF2} : for $m < n$ 
\begin{eqnarray}
\label{eq:special-ansatz}
\Psi_{0}(t)=
&&\left\{
(u_{m,11}a_{m,1}+u_{m,12}a_{m,-1})\ket{\{1, m\}}+
(u_{m,21}a_{m,1}+u_{m,22}a_{m,-1})\ket{\{-1, m\}}
\right\}+    \nonumber \\
&&\left\{
(u_{n,11}a_{n,1}+u_{n,12}a_{n,-1})\ket{\{1, n\}}+
(u_{n,21}a_{n,1}+u_{n,22}a_{n,-1})\ket{\{-1, n\}}
\right\}.
\end{eqnarray}
This ansatz is enough for our purpose because we are only interested in the 
Rabi oscillations. Then after a long calculation we have 
\begin{equation}
\label{eq:reduced-equation}
i\frac{d}{dt}
\left(
  \begin{array}{c}
    a_{m,1} \\
    a_{m,-1}
  \end{array}
\right)
=
\frac{\Delta}{2}\mbox{e}^{\Omega (m-n)}
U_{m}^{-1}
\left(
  \begin{array}{cc}
    T_{mn}& \\
      & \tilde{T}_{mn}
  \end{array}
\right)
\left(
  \begin{array}{cc}
    0& \mbox{e}^{2ig_{2}t}\\
    \mbox{e}^{-2ig_{2}t}& 0
  \end{array}
\right)
U_{n}
\left(
  \begin{array}{c}
    a_{n,1} \\
    a_{n,-1}
  \end{array}
\right),
\end{equation}
where 
\[
T_{mn}=\braa{m}\mbox{e}^{x(L_{+}-L_{-})}\kett{n}, \quad 
\tilde{T}_{mn}=\braa{m}\mbox{e}^{-x(L_{+}-L_{-})}\kett{n}
\]
and the equation with $m \longleftrightarrow n$ in 
(\ref{eq:reduced-equation}). We note $\tilde{T}_{mn}=T_{nm}$ and 
$T_{mn}=\tilde{T}_{nm}$ because $x$ is real from (\ref{eq:omega-x}). 
For the details of $T_{mn}$ or $\tilde{T}_{mn}$ see \cite{KF2}. 

We must calculate the right hand side of (\ref{eq:reduced-equation}). 
After some algebra 
\begin{eqnarray}
&&U_{m}^{-1}
\left(
  \begin{array}{cc}
    T_{mn}& \\
      & \tilde{T}_{mn}
  \end{array}
\right)
\left(
  \begin{array}{cc}
    0& \mbox{e}^{2ig_{2}t}\\
    \mbox{e}^{-2ig_{2}t}& 0
  \end{array}
\right)
U_{n}   \nonumber \\
=&&
\mbox{exp}\left\{it
\left(
  \begin{array}{cc}
    0& E_{m,\Delta} \\
    E_{m,\Delta}& -2g_{2}
  \end{array}
\right)
\right\}
\left(
  \begin{array}{cc}
    & T_{mn} \\
   \tilde{T}_{mn}& 
  \end{array}
\right)
\mbox{exp}\left\{-it
\left(
  \begin{array}{cc}
    0& E_{n,\Delta} \\
    E_{n,\Delta}& -2g_{2}
  \end{array}
\right)
\right\} \nonumber \\
=&&
\Gamma_{m}
\left(
  \begin{array}{cc}
    \mbox{e}^{it \lambda_{m,+}}& \\
      & \mbox{e}^{it \lambda_{m,-}}
  \end{array}
\right)
\Gamma_{m}^{-1}
\left(
  \begin{array}{cc}
    & T_{mn} \\
    \tilde{T}_{mn}& 
  \end{array}
\right)
\Gamma_{n}
\left(
  \begin{array}{cc}
    \mbox{e}^{-it \lambda_{n,+}}& \\
      & \mbox{e}^{-it \lambda_{n,-}}
  \end{array}
\right)
\Gamma_{n}^{-1}
\end{eqnarray}
from (\ref{eq:special-unitary}) and Appendix. Therefore the RHS of 
(\ref{eq:reduced-equation}) is 
\begin{eqnarray}
\label{eq:reduced-reduced-equation}
&&\frac{\Delta}{2}\mbox{e}^{it\Omega (m-n)}
\Gamma_{m}
\left(
  \begin{array}{cc}
    \mbox{e}^{it \lambda_{m,+}}& \\
      & \mbox{e}^{it \lambda_{m,-}}
  \end{array}
\right)
\Gamma_{m}^{-1}
\left(
  \begin{array}{cc}
    & T_{mn} \\
    \tilde{T}_{mn}& 
  \end{array}
\right)
\Gamma_{n}
\left(
  \begin{array}{cc}
    \mbox{e}^{-it \lambda_{n,+}}& \\
      & \mbox{e}^{-it \lambda_{n,-}}
  \end{array}
\right)
\Gamma_{n}^{-1},     \nonumber \\ 
{}&& 
\end{eqnarray}
where from Appendix 
\[
\lambda_{k,+}=-g_{2}+\sqrt{E_{k,\Delta}^2+g_{2}^2},\quad 
\lambda_{k,-}=-g_{2}-\sqrt{E_{k,\Delta}^2+g_{2}^2}. 
\]

From now we divide into the four cases. 
\par \noindent
Case (I) : (\ref{eq:reduced-reduced-equation}) can be written as 
\begin{eqnarray}
\label{eq:reduced-reduced-equation-I}
&&\frac{\Delta}{2}
\mbox{e}^{ it\left\{\Omega (m-n)+\lambda_{m,+}-\lambda_{n,+}\right\} }\times 
\nonumber \\
&&\Gamma_{m}
\left(
  \begin{array}{cc}
     1& \\
      & \mbox{e}^{it(\lambda_{m,-}-\lambda_{m,+})}
  \end{array}
\right)
\Gamma_{m}^{-1}
\left(
  \begin{array}{cc}
    & T_{mn} \\
    \tilde{T}_{mn}& 
  \end{array}
\right)
\Gamma_{n}
\left(
  \begin{array}{cc}
     1& \\
      & \mbox{e}^{-it(\lambda_{n,-}-\lambda_{n,+})}
  \end{array}
\right)
\Gamma_{n}^{-1}.     \nonumber 
\end{eqnarray}
Here we set {\bf the resonance condition}, namely 
\begin{equation}
\label{eq:resonance-I}
{\Omega}(m-n)+\lambda_{m,+}-\lambda_{n,+}=0, 
\end{equation}
or more explicitly 
\[
{\Omega}(m-n)+\sqrt{E_{m,\Delta}^2+g_{2}^2}-\sqrt{E_{n,\Delta}^2+g_{2}^2}=0.
\]
This is a rather complicated equation. Then the above matrix becomes 
\[
\frac{\Delta}{2}
\Gamma_{m}
\left(
  \begin{array}{cc}
     1& \\
      & \mbox{e}^{it(\lambda_{m,-}-\lambda_{m,+})}
  \end{array}
\right)
\Gamma_{m}^{-1}
\left(
  \begin{array}{cc}
    & T_{mn} \\
    \tilde{T}_{mn}& 
  \end{array}
\right)
\Gamma_{n}
\left(
  \begin{array}{cc}
     1& \\
      & \mbox{e}^{-it(\lambda_{n,-}-\lambda_{n,+})}
  \end{array}
\right)
\Gamma_{n}^{-1}.  
\]
By the way, since 
\[
\lambda_{k,-}-\lambda_{k,+}=-2\sqrt{E_{k,\Delta}^2+g_{2}^2}\quad 
\mbox{for}\quad k=m,\ n, 
\]
we can take $0$ if $g_{2}$ is large enough (so--called 
{\bf the rotating wave approximation}\footnote{in many papers in Quantum 
Optics this assumption has been used without showing that the frequency 
in the model is large enough. However it is not correct.} 
which neglects fast oscillating terms). 
Therefore we have the time--independent (!) matrix 
\begin{eqnarray}
&&\frac{\Delta}{2}
\Gamma_{m}
\left(
  \begin{array}{cc}
     1& \\
      & 0
  \end{array}
\right)
\Gamma_{m}^{-1}
\left(
  \begin{array}{cc}
    & T_{mn} \\
    \tilde{T}_{mn}& 
  \end{array}
\right)
\Gamma_{n}
\left(
  \begin{array}{cc}
     1& \\
      & 0
  \end{array}
\right)
\Gamma_{n}^{-1}  \nonumber \\
=&&\frac{\Delta}{2}
\Gamma_{m}
\left(
  \begin{array}{cc}
     \gamma_{mn}& \\
      & 0
  \end{array}
\right)
\Gamma_{n}^{-1},  \nonumber 
\end{eqnarray}
where 
\begin{eqnarray}
\gamma_{mn}
&=&
\frac{\lambda_{m,+}}{\sqrt{\lambda_{m,+}^2+E_{m,\Delta}^2}}
\tilde{T}_{mn}
\frac{E_{n,\Delta}}{\sqrt{\lambda_{n,+}^2+E_{n,\Delta}^2}}
+
\frac{E_{m,\Delta}}{\sqrt{\lambda_{m,+}^2+E_{m,\Delta}^2}}
T_{mn}
\frac{\lambda_{n,+}}{\sqrt{\lambda_{n,+}^2+E_{n,\Delta}^2}} 
\nonumber \\
&=&
\frac{1}{\sqrt{\lambda_{n,+}^2+E_{n,\Delta}^2}}
\frac{1}{\sqrt{\lambda_{m,+}^2+E_{m,\Delta}^2}}
\left(
E_{n,\Delta}T_{nm}\lambda_{m,+}
+
E_{m,\Delta}T_{mn}\lambda_{n,+}
\right) 
\end{eqnarray}
from $\tilde{T}_{mn}=T_{nm}$. 
Therefore ${\gamma}_{mn}={\gamma}_{nm}\equiv \gamma\in {\mathbf R}$. 
As a result we obtain the equations
\begin{eqnarray}
\label{eq:final-equation}
i\frac{d}{dt}
\left(
  \begin{array}{c}
    a_{m,1} \\
    a_{m,-1}
  \end{array}
\right)
&=&
\frac{\Delta}{2}\gamma
\Gamma_{m}
\left(
  \begin{array}{cc}
     1& \\
      & 0
  \end{array}
\right)
\Gamma_{n}^{-1}
\left(
  \begin{array}{c}
    a_{n,1} \\
    a_{n,-1}
  \end{array}
\right)
=
\frac{\Delta}{2}\gamma
\Gamma_{m}\frac{1_{2}+\sigma_{3}}{2}\Gamma_{n}^{-1}
\left(
  \begin{array}{c}
    a_{n,1} \\
    a_{n,-1}
  \end{array}
\right),       \nonumber  \\
{}&&                      \\
i\frac{d}{dt}
\left(
  \begin{array}{c}
    a_{n,1} \\
    a_{n,-1}
  \end{array}
\right)           
&=&
\frac{\Delta}{2}\gamma
\Gamma_{n}
\left(
  \begin{array}{cc}
     1& \\
      & 0
  \end{array}
\right)
\Gamma_{m}^{-1}
\left(
  \begin{array}{c}
    a_{m,1}  \\
    a_{m,-1}
  \end{array}
\right)
=
\frac{\Delta}{2}\gamma
\Gamma_{n}\frac{1_{2}+\sigma_{3}}{2}\Gamma_{m}^{-1}
\left(
  \begin{array}{c}
    a_{m,1}  \\
    a_{m,-1}
  \end{array}
\right),        \nonumber 
\end{eqnarray}
or if we introduce the compact notation \ 
$
{\bf a}_{m}=(a_{m,1},\ a_{m,-1})^{t}
$ 
\ then we have the clear matrix equation
\begin{equation}
i\frac{d}{dt}
\left(
  \begin{array}{c}
    {\bf a}_{m} \\
    {\bf a}_{n}
  \end{array}
\right)
=\frac{\Delta}{2}
\left(
  \begin{array}{cc}
    \Gamma_{m}&  \\
       &\Gamma_{n}
  \end{array}
\right)
\left(
  \begin{array}{cc}
    0_{2} & \gamma\frac{1_{2}+\sigma_{3}}{2} \\
    \gamma\frac{1_{2}+\sigma_{3}}{2}& 0_{2}
  \end{array}
\right)
\left(
  \begin{array}{cc}
    \Gamma_{m}&  \\
       &\Gamma_{n}
  \end{array}
\right)^{-1}
\left(
  \begin{array}{c}
    {\bf a}_{m} \\
    {\bf a}_{n}
  \end{array}
\right).
\end{equation}
Therefore the solution that we are looking for is 
\begin{eqnarray}
&&\left(
  \begin{array}{c}
    {\bf a}_{m}(t) \\
    {\bf a}_{n}(t)
  \end{array}
\right)   \nonumber \\
=&&
\left(
  \begin{array}{cc}
    \Gamma_{m}&  \\
       &\Gamma_{n}
  \end{array}
\right)
\mbox{exp}\left\{
-it\frac{\Delta}{2}
\left(
  \begin{array}{cc}
    0_{2}& \gamma\frac{1_{2}+\sigma_{3}}{2} \\
    \gamma\frac{1_{2}+\sigma_{3}}{2}& 0_{2}
  \end{array}
\right)
          \right\}
\left(
  \begin{array}{cc}
    \Gamma_{m}&  \\
       &\Gamma_{n}
  \end{array}
\right)^{-1}
\left(
  \begin{array}{c}
    {\bf a}_{m}(0) \\
    {\bf a}_{n}(0)
  \end{array}
\right). 
\end{eqnarray}
The Rabi frequency is just $\Delta \gamma$ (compare with \cite{KF2}), which 
is also rather complicated. 

Next we consider the remaining three cases : The arguments are almost 
identical, so we only give the results (we leave them to the readers).

\par \vspace{5mm} \noindent
Case (II) : {\bf The resonance condition} is 
\begin{equation}
\label{eq:resonance-II}
{\Omega}(m-n)+\lambda_{m,-}-\lambda_{n,-}=0.
\end{equation}
The solution that we are looking for is 
\begin{eqnarray}
&&\left(
  \begin{array}{c}
    {\bf a}_{m}(t) \\
    {\bf a}_{n}(t)
  \end{array}
\right)   \nonumber \\
=&&
\left(
  \begin{array}{cc}
    \Gamma_{m}&  \\
       &\Gamma_{n}
  \end{array}
\right)
\mbox{exp}\left\{
-it\frac{\Delta}{2}
\left(
  \begin{array}{cc}
    0_{2}& \gamma\frac{1_{2}-\sigma_{3}}{2} \\
    \gamma\frac{1_{2}-\sigma_{3}}{2}& 0_{2}
  \end{array}
\right)
          \right\}
\left(
  \begin{array}{cc}
    \Gamma_{m}&  \\
       &\Gamma_{n}
  \end{array}
\right)^{-1}
\left(
  \begin{array}{c}
    {\bf a}_{m}(0) \\
    {\bf a}_{n}(0)
  \end{array}
\right), 
\end{eqnarray}
where $\gamma$ is 
\begin{equation}
\gamma
=
\frac{1}{\sqrt{\lambda_{n,-}^2+E_{n,\Delta}^2}}
\frac{1}{\sqrt{\lambda_{m,-}^2+E_{m,\Delta}^2}}
\left(
E_{n,\Delta}T_{nm}\lambda_{m,-}+E_{m,\Delta}T_{mn}\lambda_{n,-}
\right). 
\end{equation}
The Rabi frequency is $\Delta \gamma$ (compare with \cite{KF2}).

\par \vspace{5mm} \noindent
Case (III) : {\bf The resonance condition} is 
\begin{equation}
\label{eq:resonance-III}
{\Omega}(m-n)+\lambda_{m,+}-\lambda_{n,-}=0.
\end{equation}
The solution that we are looking for is 
\begin{eqnarray}
&&\left(
  \begin{array}{c}
    {\bf a}_{m}(t) \\
    {\bf a}_{n}(t)
  \end{array}
\right)   \nonumber \\
=&&
\left(
  \begin{array}{cc}
    \Gamma_{m}&  \\
       &\Gamma_{n}
  \end{array}
\right)
\mbox{exp}\left\{
-it\frac{\Delta}{2}
\left(
  \begin{array}{cc}
    0_{2}& \gamma\sigma_{+} \\
    \gamma\sigma_{-}& 0_{2}
  \end{array}
\right)
          \right\}
\left(
  \begin{array}{cc}
    \Gamma_{m}&  \\
       &\Gamma_{n}
  \end{array}
\right)^{-1}
\left(
  \begin{array}{c}
    {\bf a}_{m}(0) \\
    {\bf a}_{n}(0)
  \end{array}
\right), 
\end{eqnarray}
where $\gamma$ is 
\begin{equation}
\gamma
=
\frac{1}{\sqrt{\lambda_{n,-}^2+E_{n,\Delta}^2}}
\frac{1}{\sqrt{\lambda_{m,+}^2+E_{m,\Delta}^2}}
\left(
E_{n,\Delta}T_{nm}\lambda_{m,+}+E_{m,\Delta}T_{mn}\lambda_{n,-}
\right).
\end{equation}
The Rabi frequency is $\Delta \gamma$ (compare with \cite{KF2}).

\par \vspace{5mm} \noindent
Case (IV) : {\bf The resonance condition} is 
\begin{equation}
\label{eq:resonance-IV}
{\Omega}(m-n)+\lambda_{m,-}-\lambda_{n,+}=0.
\end{equation}
The solution that we are looking for is 
\begin{eqnarray}
&&\left(
  \begin{array}{c}
    {\bf a}_{m}(t) \\
    {\bf a}_{n}(t)
  \end{array}
\right)   \nonumber \\
=&&
\left(
  \begin{array}{cc}
    \Gamma_{m}&  \\
       &\Gamma_{n}
  \end{array}
\right)
\mbox{exp}\left\{
-it\frac{\Delta}{2}
\left(
  \begin{array}{cc}
    0_{2}& \gamma\sigma_{-} \\
    \gamma\sigma_{+}& 0_{2}
  \end{array}
\right)
          \right\}
\left(
  \begin{array}{cc}
    \Gamma_{m}&  \\
       &\Gamma_{n}
  \end{array}
\right)^{-1}
\left(
  \begin{array}{c}
    {\bf a}_{m}(0) \\
    {\bf a}_{n}(0)
  \end{array}
\right), 
\end{eqnarray}
where $\gamma$ is 
\begin{equation}
\gamma
=
\frac{1}{\sqrt{\lambda_{n,+}^2+E_{n,\Delta}^2}}
\frac{1}{\sqrt{\lambda_{m,-}^2+E_{m,\Delta}^2}}
\left(
E_{n,\Delta}T_{nm}\lambda_{m,-}+E_{m,\Delta}T_{mn}\lambda_{n,+}
\right).
\end{equation}
The Rabi frequency is $\Delta \gamma$ (compare with \cite{KF2}).

\par \vspace{5mm}
On the ansatz (\ref{eq:special-ansatz}) 
we solved the Schr{\"o}dinger equation (\ref{eq:full-equation}) in the 
strong coupling regime (!) under the resonance conditions and rotating wave 
approximations, and obtained the unitary transformations of four types 
which are a generalization of \cite{KF2}. They will play a crucial role 
in Quantum Computation. 

On the other hand we in this paper solved the special case 
(\ref{eq:time-depend-F-special}), however we would like to study 
the general case (\ref{eq:time-depend-F}). At the present it is almost 
impossible (see Appendix). 
We will treat this case in a forthcoming paper. 

By the way, we considered one atom with two--level, so we would like to 
generalize our method to $n$ atoms (with two--level) interacting 
both the single radiation mode and external periodic fields like ($n$ atoms 
trapped in a cavity) 
\vspace{5mm} 
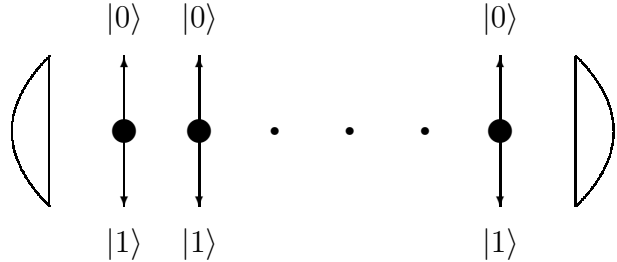
\begin{center}
\setlength{\unitlength}{1mm} 
\begin{picture}(120,40)(0,-10)
\bezier{200}(20,0)(10,10)(20,20)
\put(20,0){\line(0,1){20}}
\put(20,20){\makebox(20,10)[c]{$|0\rangle$}}
\put(30,10){\vector(0,1){10}}
\put(30,10){\vector(0,-1){10}}
\put(20,-10){\makebox(20,10)[c]{$|1\rangle$}}
\put(30,10){\circle*{3}}
\put(30,20){\makebox(20,10)[c]{$|0\rangle$}}
\put(40,10){\vector(0,1){10}}
\put(40,10){\vector(0,-1){10}}
\put(30,-10){\makebox(20,10)[c]{$|1\rangle$}}
\put(40,10){\circle*{3}}
\put(50,10){\circle*{1}}
\put(60,10){\circle*{1}}
\put(70,10){\circle*{1}}
\put(70,20){\makebox(20,10)[c]{$|0\rangle$}}
\put(80,10){\vector(0,1){10}}
\put(80,10){\vector(0,-1){10}}
\put(80,10){\circle*{3}}
\put(70,-10){\makebox(20,10)[c]{$|1\rangle$}}
\bezier{200}(90,0)(100,10)(90,20)
\put(90,0){\line(0,1){20}}
\end{picture}
\end{center}
Then the Hamiltonian may be 
\begin{equation}
\label{eq:general-n-hamiltonian}
{\tilde H}_{nL}
=\omega {\bf 1}_{M}\otimes L_{3} + 
g_{1}\sum_{j=1}^{n}\sigma_{1}^{(j)}\otimes (L_{+}+L_{-}) + 
\frac{\Delta}{2}\sum_{j=1}^{n}\sigma_{3}^{(j)}\otimes {\bf 1}_{L} + 
g_{2}\sum_{j=1}^{n}\mbox{cos}(\omega_{j}t+\phi_{j})\sigma_{1}^{(j)}\otimes 
{\bf 1}_{L}, 
\end{equation}
where $M=2^{n}$ and $\sigma_{k}^{(j)}$ ($k=1,\ 3$) is 
\[
\sigma_{k}^{(j)}=1_{2}\otimes \cdots \otimes 1_{2}\otimes \sigma_{k}\otimes 
1_{2}\otimes \cdots \otimes 1_{2}\ (j-\mbox{position}). 
\]
In the near future we will attempt an attack to this model. However according 
to increase of the number of atoms (we are expecting at least $n=100$ in the 
realistic quantum computation) we meet a very severe problem called 
Decoherence, see for example \cite{MFr3} and its references. 
The author doesn't know how to control this. 

A generalization of the model to N--level system (see for example \cite{KF3}, 
\cite{KF6}, \cite{FHKW}, \cite{KuF}) is now under consideration and will be 
published in a separate paper\footnote{The author believes that it is 
important for us to consider the N--level system to prevent the decoherence 
problem}.

\vspace{10mm}
\noindent
{\it Acknowledgment.}
The author wishes to thank Yoshinori Machida for his warm hospitality 
during the 11--th Numazu Meeting held at Numazu College of Technology 
(6--8/March/2003). 

\vspace{10mm}
\begin{center}
\begin{Large}
\noindent{\bfseries Appendix : Some Useful Formulas}
\end{Large}
\end{center}
\par \vspace{5mm} \noindent
In this appendix we solve the following equation 
\begin{equation}
i\frac{d}{dt}\psi=\alpha H\psi, 
\end{equation}
where $\alpha$ is a constant and 
\begin{equation}
H=\mbox{e}^{2i\theta t}\ket{1}\bra{-1}+\mbox{e}^{-2i\theta t}\ket{-1}\bra{1}
\quad \mbox{and}\quad \psi=a(t)\ket{1}+b(t)\ket{-1}. 
\end{equation}
Then it is easy to get a matrix equation on $\{a,\ b\}$ 
\begin{equation}
\label{eq:2-matrix-equation}
i\frac{d}{dt}
\left(
  \begin{array}{c}
    a \\
    b
  \end{array}
\right)
=\alpha
\left(
  \begin{array}{cc}
    0& \mbox{e}^{2i\theta t} \\
    \mbox{e}^{-2i\theta t}& 0
  \end{array}
\right)
\left(
  \begin{array}{c}
    a \\
    b
  \end{array}
\right)\Longleftrightarrow 
i\frac{d}{dt}{\tilde \psi}={\tilde H}{\tilde \psi}.
\end{equation}
The solution is easily obtained to become 
\begin{equation}
\left(
  \begin{array}{c}
    a \\
    b
  \end{array}
\right)
=U(t)
\left(
  \begin{array}{c}
    a_{0} \\
    b_{0}
  \end{array}
\right)
\end{equation}
where $(a_{0},b_{0})^{t}$ is a constant vector and 
\begin{equation}
U(t)=
\left(
  \begin{array}{cc}
    1& \\
     & \mbox{e}^{-2i\theta t}
  \end{array}
\right)
\mbox{exp}\left\{-it
\left(
  \begin{array}{cc}
    0& \alpha \\
    \alpha& -2\theta
  \end{array}
\right)
\right\}\Longrightarrow i\frac{d}{dt}U={\tilde H}U.
\end{equation}

\par \noindent
If we set 
\begin{equation}
U(t)=
\left(
  \begin{array}{cc}
    u_{11}& u_{12}\\
    u_{21}& u_{22}
  \end{array}
\right)
\end{equation}
($2$ corresponds to $-1$) then $\psi$ above can be written as 
\begin{equation}
\psi=(u_{11}a_{0}+u_{12}b_{0})\ket{1}+(u_{21}a_{0}+u_{22}b_{0})\ket{-1}
\end{equation}
with constants $\{a_{0},\ b_{0}\}$. 

\par \noindent 
In the method of constant variation in the text we change like 
$a_{0}\longrightarrow a_{0}(t)$ and $b_{0}\longrightarrow b_{0}(t)$. 

Let us make some comments. 
For 
\begin{equation}
A=
\left(
  \begin{array}{cc}
    0& \alpha\\
    \alpha& -2\theta
  \end{array}
\right)
\end{equation}
we can easily diagonalize $A$ as follows : 
\begin{equation}
A=
\left(
  \begin{array}{cc}
    \frac{\alpha}{\sqrt{\alpha^2+\lambda_{+}^2}}& 
    \frac{\alpha}{\sqrt{\alpha^2+\lambda_{-}^2}}\\
    \frac{\lambda_{+}}{\sqrt{\alpha^2+\lambda_{+}^2}}& 
    \frac{\lambda_{-}}{\sqrt{\alpha^2+\lambda_{-}^2}}
  \end{array}
\right)
\left(
  \begin{array}{cc}
    \lambda_{+}& \\
      & \lambda_{-}
  \end{array}
\right)
\left(
  \begin{array}{cc}
    \frac{\alpha}{\sqrt{\alpha^2+\lambda_{+}^2}}& 
    \frac{\alpha}{\sqrt{\alpha^2+\lambda_{-}^2}}\\
    \frac{\lambda_{+}}{\sqrt{\alpha^2+\lambda_{+}^2}}& 
    \frac{\lambda_{-}}{\sqrt{\alpha^2+\lambda_{-}^2}}
  \end{array}
\right)^{-1}
\end{equation}
where 
\[
\lambda_{+}=-\theta+\sqrt{\theta^{2}+\alpha^{2}},\quad 
\lambda_{-}=-\theta-\sqrt{\theta^{2}+\alpha^{2}}.
\]
Therefore we obtain 
\begin{equation}
V(t)\equiv \mbox{e}^{-itA}= 
\left(
  \begin{array}{cc}
    \frac{\alpha}{\sqrt{\alpha^2+\lambda_{+}^2}}& 
    \frac{\alpha}{\sqrt{\alpha^2+\lambda_{-}^2}}\\
    \frac{\lambda_{+}}{\sqrt{\alpha^2+\lambda_{+}^2}}& 
    \frac{\lambda_{-}}{\sqrt{\alpha^2+\lambda_{-}^2}}
  \end{array}
\right)
\left(
  \begin{array}{cc}
    \mbox{e}^{-it \lambda_{+}}& \\
      & \mbox{e}^{-it \lambda_{-}}
  \end{array}
\right)
\left(
  \begin{array}{cc}
    \frac{\alpha}{\sqrt{\alpha^2+\lambda_{+}^2}}& 
    \frac{\alpha}{\sqrt{\alpha^2+\lambda_{-}^2}}\\
    \frac{\lambda_{+}}{\sqrt{\alpha^2+\lambda_{+}^2}}& 
    \frac{\lambda_{-}}{\sqrt{\alpha^2+\lambda_{-}^2}}
  \end{array}
\right)^{-1}.
\end{equation}

\par \noindent
We note the formula in the text. 
\begin{equation}
\left(
  \begin{array}{cc}
    \frac{\alpha}{\sqrt{\alpha^2+\lambda_{+}^2}}& 
    \frac{\alpha}{\sqrt{\alpha^2+\lambda_{-}^2}}\\
    \frac{\lambda_{+}}{\sqrt{\alpha^2+\lambda_{+}^2}}& 
    \frac{\lambda_{-}}{\sqrt{\alpha^2+\lambda_{-}^2}}
  \end{array}
\right)
\left(
  \begin{array}{cc}
    1& \\
      & 0
  \end{array}
\right)
\left(
  \begin{array}{cc}
    \frac{\alpha}{\sqrt{\alpha^2+\lambda_{+}^2}}& 
    \frac{\alpha}{\sqrt{\alpha^2+\lambda_{-}^2}}\\
    \frac{\lambda_{+}}{\sqrt{\alpha^2+\lambda_{+}^2}}& 
    \frac{\lambda_{-}}{\sqrt{\alpha^2+\lambda_{-}^2}}
  \end{array}
\right)^{-1}
=
\frac{1}{\lambda_{-}-\lambda_{+}}
\left(
  \begin{array}{cc}
    \lambda_{-}& -\alpha\\
     -\alpha& -\lambda_{+}
  \end{array}
\right).
\end{equation}
For the simplicity we set 
\begin{equation}
\Gamma = 
\left(
  \begin{array}{cc}
    \frac{\alpha}{\sqrt{\alpha^2+\lambda_{+}^2}}& 
    \frac{\alpha}{\sqrt{\alpha^2+\lambda_{-}^2}}\\
    \frac{\lambda_{+}}{\sqrt{\alpha^2+\lambda_{+}^2}}& 
    \frac{\lambda_{-}}{\sqrt{\alpha^2+\lambda_{-}^2}}
  \end{array}
\right) \quad \in \quad O(2).
\end{equation}
In the text this is used as 
\begin{equation}
\Gamma_{k} = 
\left(
  \begin{array}{cc}
    \frac{E_{k,\Delta}}{\sqrt{E_{k,\Delta}^2+\lambda_{k,+}^2}}& 
    \frac{E_{k,\Delta}}{\sqrt{E_{k,\Delta}^2+\lambda_{k,-}^2}}\\
    \frac{\lambda_{k,+}}{\sqrt{E_{k,\Delta}^2+\lambda_{k,+}^2}}& 
    \frac{\lambda_{k,-}}{\sqrt{E_{k,\Delta}^2+\lambda_{k,-}^2}}
  \end{array}
\right)
\end{equation}
with 
\[
\lambda_{k,+}=-g_{2}+\sqrt{g_{2}^{2}+E_{k,\Delta}^{2}},\quad 
\lambda_{k,-}=-g_{2}-\sqrt{g_{2}^{2}+E_{k,\Delta}^{2}} 
\]
for $k=m,\ n$. 

Last we make a comment. The fact is that we wanted to solve the following 
matrix--equation instead of (\ref{eq:2-matrix-equation}) 
\begin{equation}
i\frac{d}{dt}
\left(
  \begin{array}{c}
    a \\
    b
  \end{array}
\right)
=\alpha
\left(
  \begin{array}{cc}
    0& \mbox{e}^{2i\Theta(t)} \\
    \mbox{e}^{-2i\Theta(t)}& 0
  \end{array}
\right)
\left(
  \begin{array}{c}
    a \\
    b
  \end{array}
\right), 
\end{equation}
where 
\[
\Theta(t)=\theta\frac{\mbox{sin}(\omega t)}{\omega}
={\theta t}\frac{\mbox{sin}(\omega t)}{\omega t}\ 
(\longrightarrow {\theta t}\ \ \mbox{as}\ \omega\ \rightarrow \ 0).
\]
We here note 
\begin{equation}
\mbox{e}^{2i\Theta(t)}=\sum_{n\in \futon}J_{n}(\theta/\omega)
\mbox{e}^{ni\omega t},
\end{equation}
where $J_{n}(x)$ are the Bessel functions. 

However we cannot solve this equation completely, see \cite{MFr1}, 
\cite{BaWr}, \cite{SGD}, \cite{KF5}. This is just the bottleneck.

\newpage


\end{document}